\documentclass[twocolumn,showpacs,preprintnumbers,amsmath,amssymb,pra]{revtex4}

\usepackage{graphicx}
\usepackage{dcolumn}
\usepackage{bm}

\newcommand{\ket}[1]{ { \left| #1 \right> }}

\begin{document}


\title{Simulations of the Dipole-Dipole Interaction between Two Spatially Separated Groups of Rydberg Atoms}

\author{Thomas J. Carroll}
 \affiliation{Department of Physics and Astronomy, Ursinus College, Collegeville, PA 19426.}
\author{Michael W. Noel}%
\affiliation{Physics Department, Bryn Mawr College, Bryn Mawr, PA 19010.}
\author{Christopher Daniel}
\author{Leah Hoover}
\author{Timothy Sidie}
\affiliation{Department of Physics and Astronomy, Ursinus College, Collegeville, PA 19426.}

\date{\today}

\begin{abstract}
The dipole-dipole interaction among ultra-cold Rydberg atoms is simulated.  We examine a general interaction scheme in which two atoms excited to the $x$ and $x'$ states are converted to $y$ and $y'$ states via a F\"{o}rster resonance.  The atoms are arranged in two spatially separated groups, each consisting of only one species of atom.  We record the fraction of atoms excited to the $y'$ state as the distance between the two groups is varied.  With zero detuning a many-body effect that relies on always resonant interactions causes the interaction to have a finite range.  When the detuning is greater than zero, another many-body effect causes a peak in the interaction when the two groups of atoms are some distance away from each other. To obtain these results it is necessary to include multiple atoms and solve the full many-body wave function.  These simulation results are supported by recent experimental evidence.  These many-body effects, combined with appropriate spatial arrangement of the atoms, could be useful in controlling the energy exchange among the atoms.
\end{abstract}

\maketitle

\section{\label{sec:introduction}Introduction}
Mesoscopic ensembles of cold Rydberg atoms provide an ideal laboratory for exploring quantum dynamics.  In the presence of a F\"{o}rster resonance~\cite{Forster49}, the dipole-dipole interaction allows for resonant energy exchange among the Rydberg atoms.  Due to the large dipole moments of Rydberg atoms, the interactions are long-range and take place on experimentally reasonable time scales.  The interactions can be controlled in a number of ways, including manipulating the spatial arrangement of the atoms~\cite{ditz08,carroll04,carroll06}, tuning a static electric field to shift the states of the atoms into resonance, and tailoring the mixture of Rydberg states~\cite{Mourachko04,Afrousheh04}.  The potential for precision control has led to a great deal of interest in using these systems for digital and analog quantum computing~\cite{Jaksch00,Lukin01,Protsenko02,Safronova03,Jaksch06,Brion07,Saffman08}.

In order to realize the potential of these systems, it is necessary to understand their complex many-body interactions.  The energy exchange does not occur through a pairwise sum of binary interactions, but rather through the simultaneous interactions among many atoms.  These many-body effects were first revealed through a broadening of the resonant energy exchange, which could not be accounted for by simply considering two-body interactions~\cite{anderson98,mourachko98,anderson02}.  Due to the potentially large number of atoms involved, simulation has proven to be a fruitful avenue for understanding these systems~\cite{Robicheaux04,carroll06,Weidemuller07,Younge09,Shaffer09}.  Previous work has shown that it is necessary to include the full many-body wave function for as many as 9 atoms in the calculations to reproduce experimental features~\cite{carroll06,Weidemuller06,Weidemuller07,Younge09}.

\begin{figure}
\includegraphics{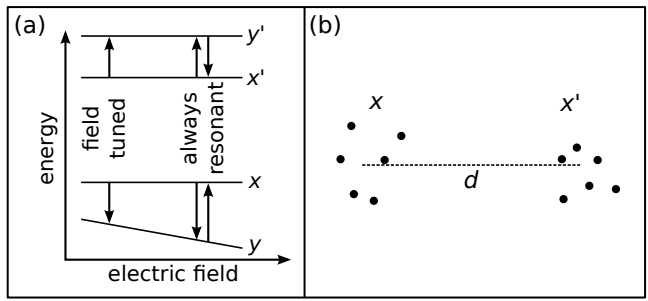}
\caption{\label{fig:LevelVolumeDiagram} (a) Energy level diagram for the resonant energy exchange of the $xx'yy'$ system. (b) The geometry studied consists of two groups of $x$ and $x'$ atoms separated by a distance $d$.}
\end{figure}

We simulate the dipole-dipole interaction among Rydberg atoms for the four state system shown in Fig.~\ref{fig:LevelVolumeDiagram}(a).  By tuning the electric field, the states can be shifted by the Stark effect such that the energy gap $E_x-E_y$ and the energy gap $E_{y'}-E_{x'}$ are made equal.  This field tuned resonant interaction is
\begin{equation}
 x + x' \rightarrow y + y',
\end{equation}
where $x \rightarrow y$ with dipole moment $\mu$ and $x' \rightarrow y'$ with dipole moment $\nu$.  There are also the always resonant interactions
\begin{eqnarray}
 x + y &\rightarrow& y + x\nonumber\\
 x' + y'&\rightarrow& y' + x'.
\end{eqnarray}
It has been suggested that the always resonant interactions contribute to an enhancement of the resonant interaction~\cite{anderson98,mourachko98}.  A system of this type has been extensively studied in Rubidium~\cite{anderson98,anderson02}, where 
\begin{eqnarray}
y&=&24p_{1/2}\nonumber\\
x&=&25s_{1/2}\nonumber\\
x'&=&33s_{1/2}\nonumber\\
y'&=&34p_{3/2}.
\end{eqnarray}
In a recent report~\cite{ditz08}, van Ditzhuijzen et al. observe the spatially resolved dipole-dipole interaction between two groups of Rydberg atoms with the energy levels
\begin{eqnarray}
y&=&49p_{3/2}\nonumber\\
x&=&49s_{1/2}\nonumber\\
x'&=&41d_{3/2}\nonumber\\
y'&=&42p_{1/2}.
\end{eqnarray}
To facilitate comparison to experiment, many of the simulations in this report are performed with parameters similar to the experimental values in Ref.~\cite{ditz08}.

The simulation is performed by diagonalization of the full dipole-dipole Hamiltonian matrix $\hat{H}$.  The Hamiltonian in atomic units is given by
\begin{multline}
 \hat{H}=\sum_{m\ne n}\hat{\rho}^m_{xy}\hat{\rho}^n_{x'y'}\frac{\mu\nu}{R_{mn}^3}+\hat{\rho}^m_{xy}\hat{\rho}^n_{yx}\frac{\mu^2}{R_{mn}^3}\\+\hat{\rho}^m_{x'y'}\hat{\rho}^n_{y'x'}\frac{\nu^2}{R_{mn}^3}+\sum_{m=n}\hat{\rho}_{yy'}^m\Delta\label{eq:Hamiltonian}
\end{multline}
where $m$ and $n$ refer to individual atoms within each state and the sum is over all atoms in each state.  The operators $\hat{\rho}_{ab}$ take an individual atom from state $a$ to state $b$, where $a$ and $b$ are the states of Fig.~\ref{fig:LevelVolumeDiagram}(a).  We ignore any orientation or spin effects and approximate the dipole-dipole interaction coupling by $\mu\nu/R_{mn}^3$ where $R_{mn}$ is the distance between the two atoms (this is similar to the Hamiltonian given in Ref.~\cite{Younge09}).  The first term in Eq.~(\ref{eq:Hamiltonian}) is the field tuned interaction and the next two terms are always resonant interactions. The final term gives the diagonal elements with a detuning, or energy defect,  $\Delta=(E_x+E_{x'})-(E_y+E_{y'})$.  While it has been found that dipole-dipole interactions can lead to consequential atomic motion~\cite{Li05,Viteau08}, we assume that we are in the regime of a ``frozen gas.''  We therefore assume that the atoms are stationary on the time scales and densities studied.  We also simplify the calculation by treating the dipole-dipole interaction as a process that occurs after the excitation of the atoms to Rydberg states, with no overlap in time.

The possible states $\ket{\phi_i}$  of the system are enumerated in the $xx'yy'$ basis. The initial state is assumed to be entirely composed of $x$ and $x'$ atoms and all excited states accessible from this initial state are included.  For monte carlo simulations, the atoms are randomly placed in two groups consisting of exclusively $x$ or $x'$ atoms and separated by a distance $d$ as shown in Fig.~\ref{fig:LevelVolumeDiagram}(b).  The results are typically averaged over hundreds of runs.  The feasibility of the monte carlo simulations is limited by how quickly the number of states scales with the total number of atoms.  For this report, we simulated up to 16 total atoms; for the case of 8 $x$ and 8 $x'$ atoms this yields 12,870 basis states.  However, we found that the results compare well to experiment when including 12 total atoms.

\section{\label{sec:SimResults}Simulation Results}
We examine the general behavior of the energy exchange between groups of Rydberg atoms by randomly placing the $x$ and $x'$ atoms in two spherical regions (see Fig.~\ref{fig:LevelVolumeDiagram}(b)).  The two spherical regions have the same radius and are separated by a distance $d$.  We simulate the interaction for some time (typically 10s of~$\mu$s in 0.1~$\mu$s steps) and the simulation results are averaged over 250 runs at each value of $d$ from 0~$\mu$m (total overlap of the two regions) to 90~$\mu$m in steps of 1~$\mu$m.  We calculate the fraction of initial $x'$ atoms found in the state $y'$, since this should be proportional to the experimental signal when using state selective field ionization.

\begin{figure}
\includegraphics{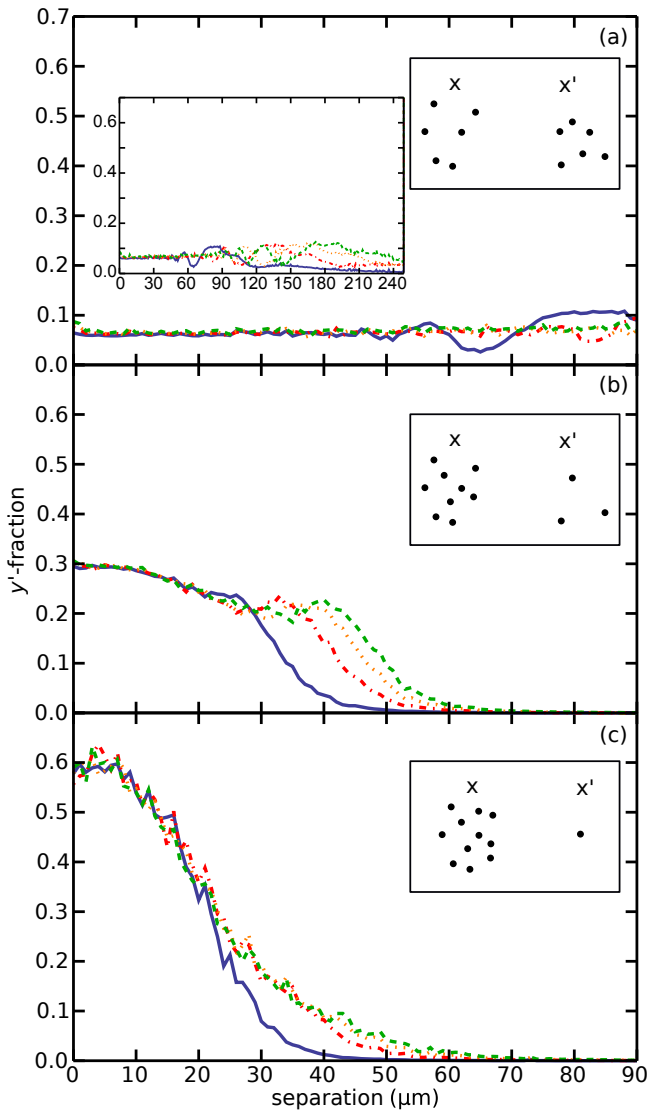}
\caption{\label{fig:12AtomSpheresZeroDelta} (color online) Fraction of initial $x'$ atoms in the $y'$ state as a function of the separation between two spherical groups of randomly placed atoms for four different times.  The detuning is zero and the total number of atoms is 12.  The times are: 1~$\mu$s (solid blue), 4~$\mu$s (dot-dashed red), 7~$\mu$s (dotted yellow), and 10~$\mu$s (dashed green). (a) When the number of $x$ and $x'$ atoms are equal (6 each), the interaction persists to large distances.  The inset shows data out to 250~$\mu$m showing that the $y'$ fraction eventually reaches zero as the Rabi period increases.  For the case of (b) 3 $x'$ atoms and 9 $x$ atoms and the case of (c) 1 $x'$ atom and 11 $x$ atoms, the fraction of atoms in the $y'$ state drops to zero with a shorter separation (around 50~$\mu$m).  }
\end{figure}

Fig.~\ref{fig:12AtomSpheresZeroDelta} shows the results for a detuning of zero when including 12 total atoms in the simulation.  Six different cases were simulated: $n$ $x'$ atoms and ($12-n$) $x$ atoms where $n=1\ldots 6$.  The dipole moments for each transition are equal, with $\mu=\nu=1000$~au.  Due to this symmetry, the fraction of atoms in the $y'$ state  for $n=7\ldots11$ can be inferred from the same data. For the case where the number of $x$ and $x'$ atoms are equal, shown in Fig.~\ref{fig:12AtomSpheresZeroDelta}(a), the energy exchange persists to large separation.  For the remaining cases, two of which are shown in Fig.~\ref{fig:12AtomSpheresZeroDelta}(b) and Fig.~\ref{fig:12AtomSpheresZeroDelta}(c), the $y'$ fraction drops to zero at smaller separation even though the detuning is zero. The finite range of the interaction in the absence of any detuning is due to a many-body effect that will be discussed in Sec.~\ref{sec:threeatom}.

\begin{figure}
\includegraphics{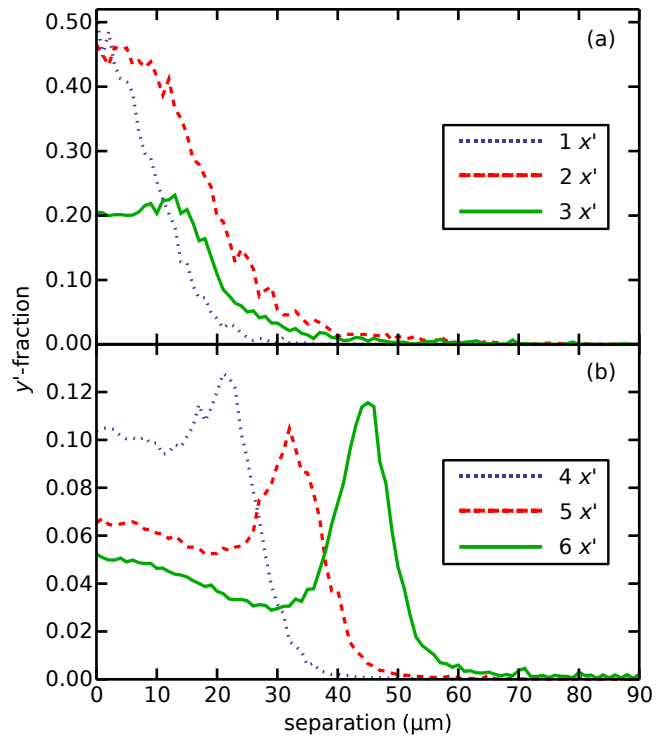}
\caption{\label{fig:12AtomSpheresPosDelta} (color online) Fraction of atoms in the $y'$ state as a function of the separation between two spherical groups of randomly placed atoms after an interaction time of 5~$\mu$s. The detuning $\Delta\approx 4$~MHz and the total number of atoms is 12.  The most prominent feature is the peak in the interaction that occurs away from overlap for configurations with 3 to 6 $x'$ atoms.}
\end{figure}

The most interesting feature appears when the detuning is greater than zero.  Fig.~\ref{fig:12AtomSpheresPosDelta} shows simulation data generated in the same way as the data in Fig.~\ref{fig:12AtomSpheresZeroDelta}, with the exception that the detuning $\Delta\approx 2$~MHz. In Fig.~\ref{fig:12AtomSpheresPosDelta} the peak in the ``strength'' of the interaction, as measured by the $y'$ fraction, is significantly away from the overlap of the two regions.  The location of the peak is persistent for all times.  In all cases shown in Fig.~\ref{fig:12AtomSpheresPosDelta}, the $y'$ fraction drops to zero with a finite range due to the combined effect of the detuning and many-body effects.

\begin{figure}
\includegraphics{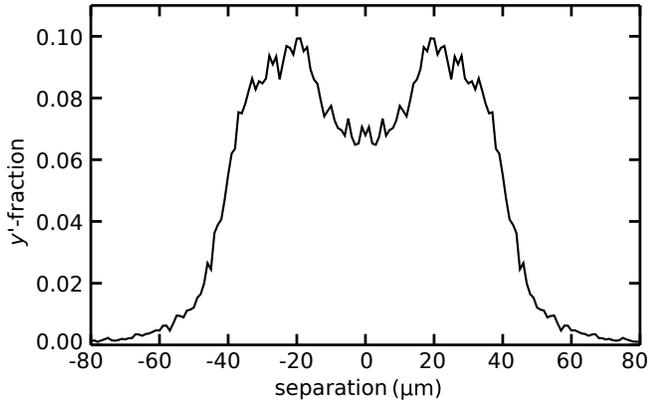}
\caption{\label{fig:ComparisonToExperiment} Fraction of atoms in the $y'$ state after an interaction time of 5~$\mu$s as a function of the separation between the two groups of Rydberg atoms with volumes defined by excitation lasers with gaussian beam profiles.  The detuning is $\Delta\approx 2$~MHz and the total number of atoms is 12.  The location of the interaction peaks away from overlap at about 20~$\mu$m and the sharp turn-off of the interaction from about 40-50~$\mu$m compare well to the data in Ref.~\cite{ditz08}.}
\end{figure}

Before exploring the detailed behavior and origin of the peak in Sec.~\ref{sec:Peak}, we note that this feature has been seen by van Ditzhuijzen et al. in their recent work demonstrating interactions between spatially resolved groups of Rydberg atoms~\cite{ditz08}.  We have run simulations roughly mimicking their experimental parameters; the results are shown in Fig.~\ref{fig:ComparisonToExperiment}.  The two regions of atoms are modeled as two gaussian beams each with a beamwaist of 14~$\mu$m and a length of 250~$\mu$m.  To account for different numbers of $x$ and $x'$ atoms in each beam, different cases for 12 total atoms that are similar to the Rydberg populations cited in the experiment are averaged.  The detuning is $\Delta\approx 2$~MHz.  With some adjustment of the detuning, similar results can be obtained for numbers of atoms larger than 12.

The data in Fig.~\ref{fig:ComparisonToExperiment} is graphed for positive and negative separations, corresponding to the beam of $x$ atoms being displaced to either side of the the $x'$ beam.  While the simulation is manifestly symmetric about the overlap of the two beams (0~$\mu$m), we graph Fig.~\ref{fig:ComparisonToExperiment} in this manner to facilitate comparison to the experimental results in Ref.~\cite{ditz08}.  Our simulation data agrees well with the experimental data, particularly on the two aformentioned prominent features.  First, the fraction of atoms excited to the $y'$ state drops rapidly to zero from about 40 to 50~$\mu$m.  Second, the fraction of atoms excited to the $y'$ state peaks at a separation between the beams of about 20~$\mu$m. However, our predicted mixing fraction is less than the observed mixing fraction.  It is possible that this is due to ignoring the temporal overlap of the Rydberg excitation process and the dipole-dipole interactions, which has been found to increase the mixing fraction~\cite{Younge09}.

\section{\label{sec:ManyBody}Theory}
\subsection{\label{sec:threeatom}Three Atom Model: One $x$ and Two~$x'$ Atoms with Zero Detuning}

\begin{figure}
\includegraphics{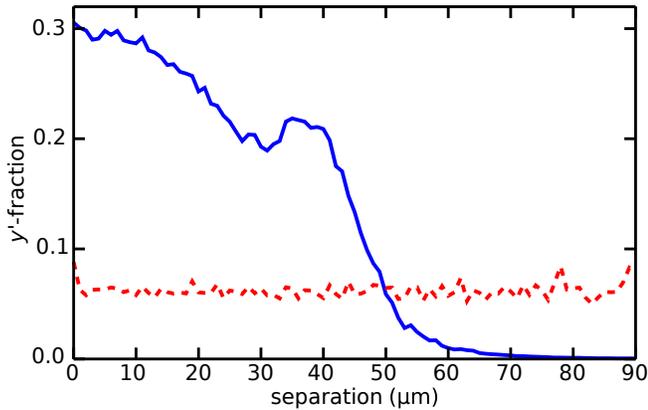}
\caption{\label{fig:aroff} Fraction of atoms in the $y'$ state as a function of the separation between two spherical groups of randomly placed atoms after an interaction time of 5~$\mu$s. In this case, there are 3 $x'$ and 9 $x$ atoms.  The solid blue line is for $\Delta$=0 and is from the same data as Fig.~\ref{fig:12AtomSpheresZeroDelta}(b).  The same simulation was run with the always resonant interactions removed and the result is shown with the dashed red line.  While the $y'$ fraction is smaller, it also persists to large separations, implicating the always resonant interactions as the cause for the finite range of the interaction at zero detuning.}
\end{figure}

When there are unequal numbers of $x$ and $x'$ atoms in the two groups, the $y'$ fraction drops to nearly zero within a finite range (see Figs.~\ref{fig:12AtomSpheresZeroDelta}(a) and~\ref{fig:12AtomSpheresZeroDelta}(b)).  Interference from always resonant interactions among atoms in each group suppresses the field tuned interaction between the two groups and leads to less population transfer to the $y'$ state.  In the simulation it is possible to remove the always resonant terms from the Hamiltonian in Eq.~\ref{eq:Hamiltonian}, effectively turning off the always resonant interactions.  The result, when applied to the case of 3 $x'$ atoms and 9 $x$ atoms (Fig.~\ref{fig:12AtomSpheresZeroDelta}(b)), is shown in Fig.~\ref{fig:aroff}.  At small separations, with the always resonant interactions active, the energy exchange is enhanced and more atoms are found in the $y'$ state.  However, at larger separations, the energy exchange is strongly suppressed.  With no always resonant interactions the $y'$ fraction, while smaller, persists to large separations.

\begin{figure}
\includegraphics{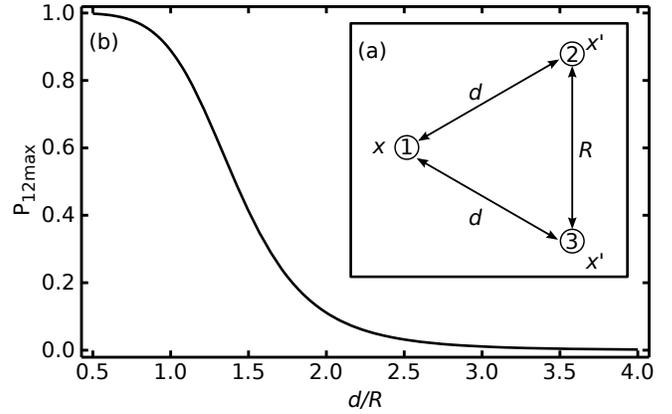}
\caption{\label{fig:threeatoms} (a) One $x$ atom at a distance $d$ from two $x'$ atoms, which are separated by a distance $R$.  We hold the distance $R$ fixed and consider what happens as $d$ is varied.  (b) The maximum probability of the three atoms being in the one of the states $\ket{\phi_{e1}}$ or $\ket{\phi_{e2}}$ vs. the separation $d$ in units of $R$.}
\end{figure}

The effect of the always resonant interactions among groups of $x$ and $x'$ atoms can be examined analytically with a simple three atom model with zero detuning.  An $x$ atom is placed a distance $d$ from each of two $x'$ atoms that are separated by a distance $R$.  This geometry is shown in Fig.~\ref{fig:threeatoms}(a), where the $x$ atom is labeled 1 and the $x'$ atoms are labeled 2 and 3.  The field tuned interaction between the $x$ atom and an $x'$ atom is given by $u=\mu\nu/d^3$.  The always resonant interaction between the two $x'$ atoms is given by $v=\nu^2/R^3$.  For simplicity, in the following analysis, we choose $\mu=\nu$.  If $\mu\ne\nu$, this will only change the distance scale of the predicted behavior.  Numerical calculations show that the following results are insensitive to the exact placement of the atoms.

In the zero interaction basis, we write the states of the atoms as
\begin{eqnarray} 
\ket{\phi_g}&=&\ket{x}_1\ket{x'}_2\ket{x'}_3\nonumber\\
\ket{\phi_{e1}}&=&\ket{y}_1\ket{y'}_2\ket{x'}_3\nonumber\\
\ket{\phi_{e2}}&=&\ket{y}_1\ket{x'}_2\ket{y'}_3\label{eqn:zerointeractionstates}
\end{eqnarray}
where the subscripts on the right hand side are the atom labels.  The initial state $\ket{\phi_g}$ is connected to the ``excited'' states $\ket{\phi_{e1}}$ and $\ket{\phi_{e2}}$ via field tuned interactions between atom 1 and either atom 2 or atom 3.  The two excited states are connected to each other via the always resonant interaction between atoms 2 and 3.

We can write the time-dependent Schr\"{o}dinger equation as
\begin{eqnarray}
 i\dot{c}_g&=&uc_{e1}+uc_{e2}\nonumber\\
i\dot{c}_{e1}&=&uc_g+vc_{e2}\nonumber\\
i\dot{c}_{e2}&=&uc_g+vc_{e1}
\end{eqnarray}
where $c_g$ is the amplitude for $\ket{\phi_g}$, $c_{e1}$ is the amplitude for $\ket{\phi_{e1}}$, and $c_{e2}$ is the amplitude for $\ket{\phi_{e2}}$.
The solutions when the atoms are initially in state $\ket{\phi_g}$ are
\begin{eqnarray}
c_g&=&\frac{1}{2\sqrt{8u^2+v^2}}((\sqrt{8u^2-v^2}-v)e^{-\frac{i}{2}(\sqrt{8u^2+v^2}+v)t}\nonumber\\
&&+(\sqrt{8u^2-v^2}+v)e^{\frac{i}{2} (\sqrt{8u^2+v^2}-v)t})\nonumber\\
c_{e1}&=&\frac{u}{2\sqrt{8u^2+v^2}}(e^{-\frac{i}{2} (\sqrt{8u^2+v^2}+v)t}-e^{\frac{i}{2}(\sqrt{8u^2+v^2}-v)t})\nonumber\\
c_{e2}&=&\frac{u}{2\sqrt{8u^2+v^2}}(e^{-\frac{i}{2}(\sqrt{8u^2+v^2}+v)t}-e^{\frac{i}{2}(\sqrt{8u^2+v^2}-v)t}).\nonumber\\\label{eqn:solns}
\end{eqnarray}
We examine the behavior of the system in terms of $P_{12}=1-|c_g|^2$, the probability of finding the atoms in one of the excited states.  From Eq.~\ref{eqn:solns}, we find
\begin{eqnarray}
 P_{12}=1-\frac{4u^2+v^2}{8u^2+v^2}+\frac{4u^2}{8u^2+v^2}\cos(\sqrt{8u^2+v^2}~t).
\end{eqnarray}
Note that by setting the always resonant interaction $v=0$, we recover standard Rabi oscillations whose only dependence on the separation $d$ is in the frequency.

The case where the $x$ atom is placed equidistant between the $x'$ atoms ($d=R/2$ or $u=8v$) yields a solution nearly identical to standard Rabi oscillations except that the maximum probability $P_{12max}$ is slightly less than one.  As we increase $d$ and move the $x$ atom farther from the pair of $x'$ atoms, the maximum probability of the system being found in one of the excited states $\ket{\phi_{e1}}$ or $\ket{\phi_{e2}}$ decreases rapidly, as shown by plotting $P_{12max}$ in Fig.~\ref{fig:threeatoms}(b).

As $d/R$ increases, the always resonant interaction increasingly suppresses the field tuned interaction, even at zero detuning.  This effect is similar to the dark states phenomenon~\cite{Stroud82}.  The dressed states of this three atom system, which are superpositions of the states of Eq.~\ref{eqn:zerointeractionstates}, depend on $d$.  As $d$ increases, one of the dressed states becomes nearly identical to $\ket{\phi_g}$.  The initial state $\ket{\phi_g}$ thus becomes dark, totally decoupled from the other states.  The evolution of the initial state to a dark state plays an important role in explaining the peak in the interaction as well.

\subsection{\label{sec:fouratom}Four Atom Model: Two $x$ and Two~$x'$ Atoms with Zero Detuning}

\begin{figure}
\includegraphics{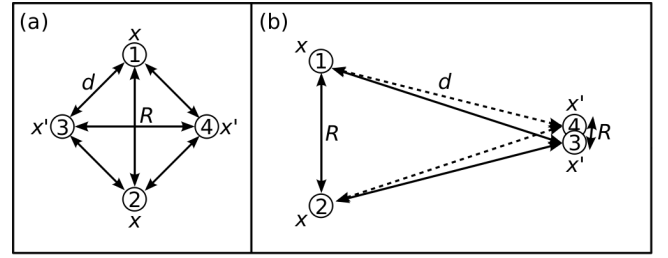}
\caption{\label{fig:fouratom} Two $x$ atoms and two $x'$ atoms arranged so that the separation between any $xx'$ pair is always $d$.  The pair of $x$ atoms and the pair of $x'$ atoms each comprise a ``group'' of atoms.  The separation within the pairs is $R$ and is held fixed.  (a) The two groups of atoms at their smallest separation; all of the atoms are in the same plane.  This view is along the axis of separation. (b) A view of the atoms from the ``side.''  From this view, atoms 3 and 4 are on top of each other and the distance $R$ between them is foreshortened.}
\end{figure}

The three atom model does not address the results of Fig.~\ref{fig:12AtomSpheresZeroDelta}(a), which show that the range of the energy exchange at zero detuning for equal numbers of $x$ and $x'$ atoms is limited only by the interaction time (see particularly the inset in Fig.~\ref{fig:12AtomSpheresZeroDelta}(a)).  We construct the four atom model shown in Fig.~\ref{fig:fouratom}(a)-(b) to examine the asymptotic behavior of the energy exchange at large separations for equal numbers of $x$ and $x'$ atoms.  Numerical calculations show that the following results are insensitive to the particular choice of geometry so we choose an arrangement that allows us to obtain an analytical solution.  Both the two $x$ atoms and the two $x'$ atoms are separated by a distance $R$.  The distance between any $x$ atom and any $x'$ atom is $d$.  As before, the field tuned interaction is given by $u=\mu\nu/d^3$ and the always resonant interaction by $v=\mu^2/d^3$.

In the zero-interaction basis, we write the states of the atoms as 
\begin{eqnarray}
 \ket{\phi_g}&=&\ket{x}_1\ket{x}_2\ket{x'}_3\ket{x'}_4\nonumber\\
\ket{\phi_{e1}}&=&\ket{y}_1\ket{x}_2\ket{y'}_3\ket{x'}_4\nonumber\\
\ket{\phi_{e2}}&=&\ket{y}_1\ket{x}_2\ket{x'}_3\ket{y'}_4\nonumber\\
\ket{\phi_{e3}}&=&\ket{x}_1\ket{y}_2\ket{y'}_3\ket{x'}_4\nonumber\\
\ket{\phi_{e4}}&=&\ket{x}_1\ket{y}_2\ket{x'}_3\ket{y'}_4\nonumber\\
\ket{\phi_{e5}}&=&\ket{y}_1\ket{y}_2\ket{y'}_3\ket{y'}_4
\end{eqnarray}
where the subscripts on the right hand side are the atom labels.  We can write the time-dependent Schr\"{o}dinger equation as
\begin{eqnarray}
i\dot{c}_g&=&uc_{e1}+uc_{e2}+uc_{e3}+uc_{e4}+4uc_{e5}\nonumber\\
i\dot{c}_{e1}&=&uc_g+vc_{e2}+vc_{e3}+2(u+v)c_{e4}+uc_{e5}\nonumber\\
i\dot{c}_{e2}&=&uc_g+vc_{e1}+2(u+v)c_{e3}+vc_{e4}+uc_{e5}\nonumber\\
i\dot{c}_{e3}&=&uc_g+vc_{e1}+2(u+v)c_{e2}+vc_{e4}+uc_{e5}\nonumber\\
i\dot{c}_{e4}&=&uc_g+2(u+v)c_{e1}+vc_{e2}+vc_{e3}+uc_{e5}\nonumber\\
i\dot{c}_{e5}&=&4uc_g+uc_{e1}+uc_{e2}+uc_{e3}+uc_{e4}
\end{eqnarray}
where $c_g$ is the amplitude for the initial state $\ket{\phi_g}$ and $c_{ei}$ is the amplitude for the excited state $\ket{\phi_{ei}}$.

If we keep only terms that are first-order in $u$, the solution for $c_g$ when the atoms are initially in state $\ket{\phi_g}$ is
\begin{equation}
c_g=\frac{1}{2} e^{4 i u t}+\frac{1}{2} e^{-4 i u t}+\frac{3 u e^{4 i u t}}{32 v}-\frac{3 u e^{-4 i u t}}{32 v}
\end{equation}
so that the probability of the system being found in the state $\ket{\phi_g}$ is
\begin{equation}
 |c_g|^2=\cos ^2(4 u t)+\frac{9 u^2 \sin ^2(4 u t)}{1024 v^2}.
\end{equation}
At large separations, the only dressed states coupled to $\ket{\phi_g}$ are equal superpositions of $\ket{\phi_g}$ and $\ket{\phi_{e5}}$.  The always resonant interactions have a neglible effect and the energy exchange resembles simple Rabi oscillations between the two groups of atoms.

\subsection{\label{sec:Peak}Interaction Peak away from Overlap}
\begin{figure}
\includegraphics{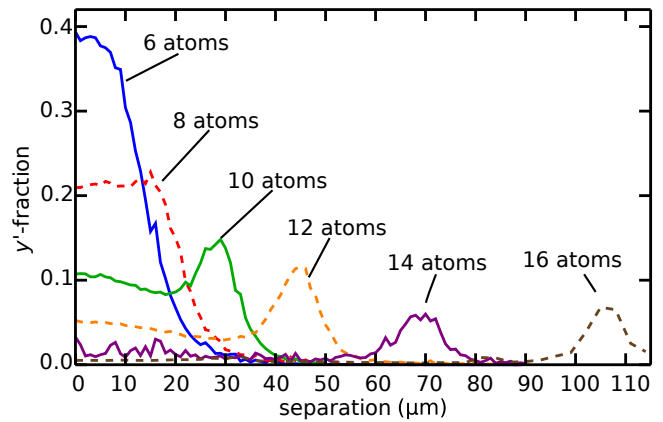}
\caption{\label{fig:PeakForDifferentNumbers} (color online) Fraction of atoms in the $y'$ state as a function of the separation between two amorphous spherical groups of atoms.  The detuning $\Delta\approx 4$~MHz.  Shown here are the cases for which the number of $x$ and $x'$ atoms are equal for different total numbers of atoms.  The peak in the interaction away from overlap moves to larger separations and becomes smaller as the total number of atoms is increased.  Unless a sufficient number of atoms is included in the calculation, the peak is not observed.}
\end{figure}

A three or four atom model is insufficient to model the enhancement in the interaction away from overlap.  This is evident if we vary the number of atoms included in the calculation.  Fig.~\ref{fig:PeakForDifferentNumbers} shows simulation data for different total numbers of atoms, from 6 to 16, with equal numbers of $x$ and $x'$ atoms in each case.  This data is generated in the same fashion as the data for Figs.~\ref{fig:12AtomSpheresZeroDelta} and~\ref{fig:12AtomSpheresPosDelta}, with the exception of the 14 and 16 atom data.  Since the number of basis states is large for these two cases (3,432 and 12,870), the 14 atom data was generated with reduced averaging (100 runs per 1~$\mu$m) and the 16 atom data was generated with reduced averaging and reduced resolution (50 runs per 3~$\mu$m).  For different total numbers of atoms the radius of the spherical regions is adjusted to hold the density constant.  The location of the peak depends strongly on the number of atoms included in the simulation.

With an amporphous sample of atoms, no peak in the interaction is observed in the 6 atom data in Fig.~\ref{fig:PeakForDifferentNumbers}.  We will see that under certain conditions a weak effect may be observed in the 6 atom case.  However, no effect is observed for 5 atoms or fewer.  The presence of the peak in the interaction away from overlap is therefore an intrinsically many-body effect, requiring at least 6 atoms to manifest.

\begin{figure}
\includegraphics{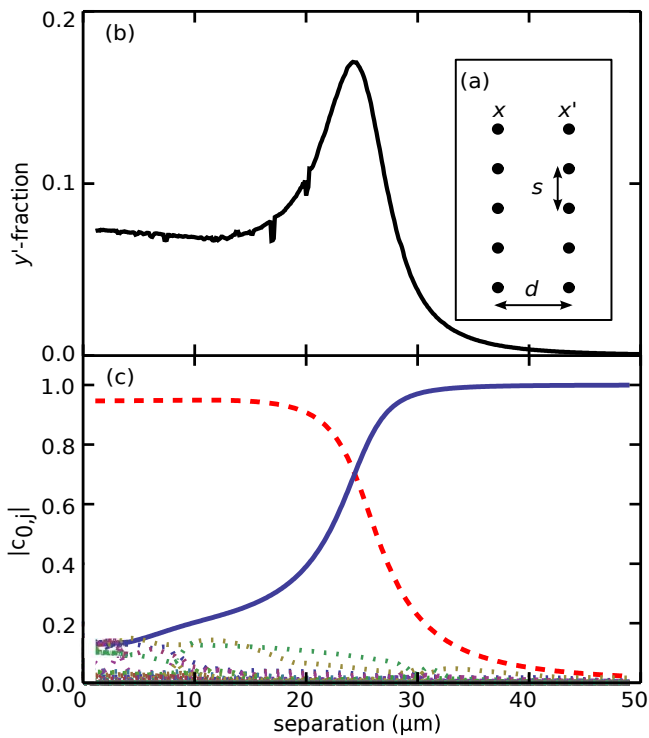}
\caption{\label{fig:FiveAtomLine} (color online) (a) Geometry for a linear array of atoms where $d$ is the distance between the lines and $s$ is the spacing between the atoms.  (b) Fraction of atoms in the $y'$ state averaged over 10~$\mu$s as a function of the separation $d$ between two lines of 5 atoms with $s$=10~$\mu$m.  The separation $d$ starts at 1~$\mu$m and is increased in steps of 0.01~$\mu$m.  As before, there is a peak in the interaction away from the overlap at about 25~$\mu$m.  (c) The coefficients $|c_{0,j}|$ linking each dressed state to the initial state as a function of position.  Only two of the 252 dressed states are significantly coupled to the initial state; they are shown in solid blue and dashed red (the remaining 250 coefficients are shown in dotted lines).  One dressed state (solid blue) is weakly coupled to the initial state at small separations but evolves to be nearly identical to the initial state at large separations.  Another dressed state (dashed red) is strongly coupled to the initial state at small separations and evolves into a state that is nearly totally decoupled from the initial state at large separations.  Where these two dressed states cross, they are both equally coupled to the initial state and coupled to many other excited states.  At this crossing, there is a strong peak in the interaction shown in (b).}
\end{figure}

\begin{figure}
\includegraphics{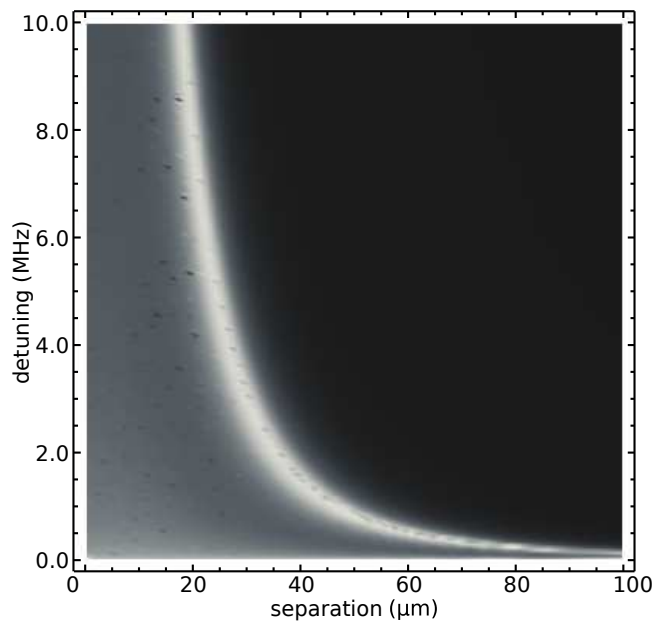}
\caption{\label{fig:manyDeltas} Fraction of atoms in the $y'$ state averaged over 10~$\mu$s as a function of the separation and the detuning.  The atoms are arranged as in Fig.~\ref{fig:FiveAtomLine}(a). Lighter shades of gray correspond to larger fractions of atoms in the $y'$ state.  The location of the peak in the interaction away from overlap can be seen as a light band that curves from the upper-left to the lower-right.  The resolution of this data is 1~$\mu$m and about 0.07~MHz.}
\end{figure}

In order to explore the many-body interactions in more detail, we examine the results for a simpler geometry of fixed instead of random atom positions.  The two groups of atoms are arranged in a regularly spaced linear array, as shown in Fig.~\ref{fig:FiveAtomLine}(a), with the atoms spaced by 10~$\mu$m.  Using this model, we can study the dependence of the peak position on the detuning. Fig.~\ref{fig:manyDeltas} is an intensity plot of the time-averaged $y'$ fraction as a function of the separation between two lines of 5 atoms and the detuning.  The detuning is varied from 0~MHz to 10~MHz, corresponding to a few orders of magnitude smaller to a few orders of magnitude larger than the dipole-dipole interaction matrix elements, depending on the separation between the lines.

As the detuning approaches zero, the peak in the interaction moves to larger separations and the interaction eventually becomes constant for all separations at $\Delta=0$. In an experiment, sources of detuning might include not only the experimental choice of electric field, but also the effects of electric field inhomogeneity and potentially a trapping magnetic field.  In our simulations, we choose $\Delta$ to achieve reasonable agreement with experiment, as in Fig.~\ref{fig:ComparisonToExperiment}; in this sense $\Delta$ should be considered a free parameter.  However, the location of the peak changes rapidly only at small $\Delta$. Once $\Delta\approx 2$~MHz the peak location changes slowly, so our results are robust for a wide range of detunings.

The presence of the peak in the interaction away from overlap can be understood by examining the dressed states of our solution.  In the non-interacting basis, we can write the states of the system as
\begin{equation}
 \ket{\phi_i}=\ket{\alpha}_1\ket{\alpha}_2\cdots\ket{\alpha}_M\ket{\beta}_{M+1}\ket{\beta}_{M+2}\cdots\ket{\beta}_N
\end{equation}
where $\alpha=x$ or $y$, $\beta=x'$ or $y'$, the number of $x$ and $y$ atoms is $M$ and the number of $x'$ and $y'$ atoms is $(N-M)$, and the subscripts on the right hand side refer to individual atoms.  The initial state $\ket{\phi_0}$ is composed of only $x$ and $x'$ atoms.  At a particular separation $d$ between the two lines, we can write the dressed states as a superposition of the $\ket{\phi_i}$ with
\begin{equation*}
 \ket{\psi_j}=\sum_i c_{i,j}\ket{\phi_i}.
\end{equation*}

The coefficient $c_{0,j}$ determines the coupling of each dressed state to the initial state $\ket{\phi_0}$.  In Fig.~\ref{fig:FiveAtomLine}(c) the coefficients $|c_{0,j}|$ are plotted as a function of the separation between the two lines of atoms for the case of 5 atoms in each line with a detuning of about 6~MHz and a spacing of 10~$\mu$m (see Fig.~\ref{fig:FiveAtomLine}(a)).  As in the three atom model, one of the dressed states (solid blue) becomes nearly identical to the initial state beyond about 30~$\mu$m.  The initial state is then dark, entirely decoupled from the other states~\cite{Stroud82}.  As shown in Fig.~\ref{fig:FiveAtomLine}(b), this corresponds to a region where the $y'$ fraction rapidly approaches zero.  However, at small separations (less than 20~$\mu$m), the eventual dark state is weakly coupled to the initial state.  Another dressed state (dashed red) is strongly coupled to the initial state at small separations and thus weakly coupled to the excited states, corresponding to a region in Fig.~\ref{fig:FiveAtomLine}(b) where the $y'$ fraction is small but non-zero.  At separations between 20 and 30~$\mu$m, these two dressed states switch roles and the coefficients $|c_{0,j}|$ cross.  In this region, both dressed states are equally strongly coupled to the initial state and well-mixed with many excited states.  This mixing yields more population transfer to the $y'$ state, creating a peak in the $y'$ fraction away from overlap in Fig.~\ref{fig:FiveAtomLine}(b).

\begin{figure}
\includegraphics{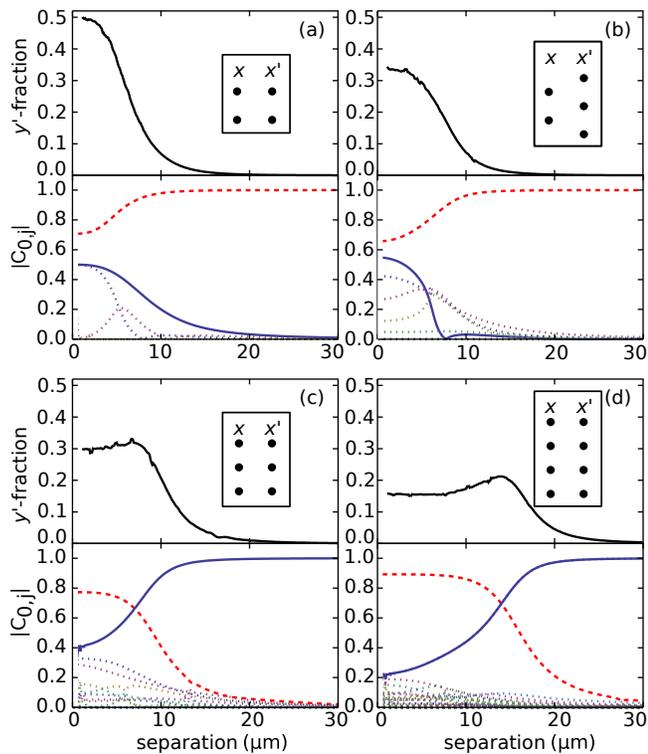}
\caption{\label{fig:Eigenvectors} (color online) Dressed state coefficients $|c_{0,j}|$ and $y'$ fractions plotted together for the cases: (a) two lines of 2 atoms, (b) one line of 2 atoms and one line of 3 atoms, (c) two lines of 3 atoms, and (d) two lines of 4 atoms.  With 5 or fewer total atoms, as in (a) and (b), the $y'$ fraction peaks at overlap.  With 6 total atoms, as in (c), a peak in the interaction away from overlap just begins to form near a 6~$\mu$m separation.  As more atoms are included, as in (d) and Fig.~\ref{fig:FiveAtomLine}, the peak becomes more pronounced and moves to larger separations.}
\end{figure}

Figs.~\ref{fig:Eigenvectors}(a)-(d) show a series of plots similar to Fig.~\ref{fig:FiveAtomLine}, including the results for 4, 5, 6, and 8 atoms respectively.  In the 4 and 5 atom cases, shown in Figs.~\ref{fig:Eigenvectors}(a) and (b), there is no crossover point.  The dressed state which evolves to become identical to the initial state is the most strongly coupled to the initial state even at small separations.  Thus, there is no peak in the interaction in Figs.~\ref{fig:Eigenvectors}(a) or (b).  When 6 or more atoms are included in the simulation, as in Figs.~\ref{fig:Eigenvectors}(c) and (d), the coefficients cross and a peak away from overlap begins to form.

\section{\label{sec:conclusion}Conclusion}
Our results show, in agreement with other recent work, that in order to correctly model the collective interactions among Rydberg atoms it is necessary to calculate the full many-body wave function.  In  Refs.~\cite{Weidemuller07} and~\cite{Younge09}, the authors conclude that at least 4 or 5 atoms must be included to accurately model their experiment and that summing over binary interactions is not sufficient.  Here, we find good agreement using 12 atoms and require a minimum of 6 atoms to observe the interaction peak away from overlap.  However, we note that while the full many-body wave function is necessary to accurately model experiment, the results in Figs.~\ref{fig:FiveAtomLine} and~\ref{fig:Eigenvectors} suggest that only two of the dressed states play a dominant role.  Given the number of atoms and possible states involved in the simulation, this is indicative of the collective nature of the interactions.  It also suggests that some simplification of the analysis may be possible by considering the atoms collectively.

While it is clear that precisely positioning the Rydberg atoms will yield some degree of control over their interactions, our results show that that significant control may be possible even with amorphous samples.  The peak in the interaction away from overlap offers a one such avenue for experimental control. In Fig.~\ref{fig:manyDeltas} there is a large region of parameter space where the interaction is nearly zero.  Separating two groups of atoms and adjusting the detuning so that there is no interaction, one could place a system in this region.  By slightly adjusting the detuning, one could then switch from no interaction to a relatively strong level of interaction.  The interaction could be similarly controlled by changing the position of one or both groups of atoms.

This material is based upon work supported by the National Science Foundation under Grant No. 0653544.

\bibliography{carroll09}

\end{document}